\documentclass[twocolumn,showpacs,pre,floatfix]{revtex4-1}
\usepackage{amsfonts}
\usepackage{amssymb}
\usepackage{amsmath}
\usepackage{graphicx}

\usepackage[latin1]{inputenc}

\newcommand{\ic}{\mathrm{i}}

\begin{document}

\title{Trace formula for dielectric cavities  III: TE modes}
\author{E. Bogomolny}
\affiliation{Univ. Paris Sud, CNRS,  LPTMS, UMR 8656,  Orsay F-91405, France}
\author{R. Dubertrand}
\affiliation{School of Mathematics, University Walk, Bristol BS8
  1TW, United Kingdom} 

\date{\today}

\begin{abstract}
The  construction of the semiclassical trace formula for the
resonances with the transverse electric (TE)  polarization for
two-dimensional dielectric
cavities is  discussed. Special attention is given to the derivation of the
two first terms of
Weyl's series for the average number of such resonances.  The obtained
formulas
agree well with numerical calculations for dielectric cavities of different
shapes. 
\end{abstract}
\maketitle


\section{Introduction}\label{intro}

Open dielectric cavities  have attracted  a large interest in recent years 
due to their numerous and potentially important applications
\cite{vahala,matsko}. From a theoretical point of view, the crucial
difference between dielectric cavities and much more investigated case of
closed  quantum billiards \cite{balian_bloch,gutzwiller,haake} is that in
the latter the spectrum is discrete but in the former it is continuous. 
Indeed, the main subject of investigations in open systems is not the true
spectrum but the spectrum of resonances defined as poles of the scattering
$S$-matrix (see e.g.  \cite{scattering,s_matrix}). 

The wavelength of electromagnetic field is usually much smaller than
any characteristic cavity size (except its height) and semiclassical
techniques are useful and adequate for a theoretical approach to
such objects. It is well known that the trace formulas are a very powerful
tool in
the semiclassical  description  of closed  systems, see e.g.
\cite{balian_bloch,gutzwiller,haake}. Therefore, the
generalization of trace formulas to different open systems, in particular to
dielectric cavities,  is of importance.  

The trace formula for resonances with transverse magnetic (TM) polarization 
in two-dimensional
(2d) dielectric cavities has been developed in
\cite{trace1} and shown to agree well with the
experiments and numerical calculations \cite{trace2,stefan2010-2012}.  

This paper is devoted to the construction of  the trace formula for 2d
dielectric
cavities but for transverse electric (TE) polarization. Due to different
boundary conditions the case of TE modes differs in many aspects from TM
modes. In particular, a special treatment is required for the resonances
related
to Brewster's angle \cite{jackson} at which the Fresnel reflection
coefficient
vanishes. 

Our main result is the asymptotic formula in the semiclassical (aka short
wave
length) regime for the average number of
TE resonances for a 2d dielectric cavity  with refraction index $n$, area
$\mathcal{A}$ and perimeter $\mathcal{L}$ 
\begin{equation}
\bar{N}_{\mathrm{TE}}(k)=\frac{\mathcal{A}\, n^2
k^2}{4\pi}+r_{\mathrm{TE}}(n)\frac{\mathcal{L}\, k}{4\pi}+o(k)\ .
\end{equation}
Here $\bar{N}_{\mathrm{TE}}(k)$ is the mean number of resonances (defined
below) whose real part is less than $k$, the coefficient $r_{\mathrm{TE}}$ is
given by the  expression 
\begin{eqnarray}
r_{\mathrm{TE}}(n)&=&1+\frac{1}{\pi}\int_{-\infty}^{\infty}\tilde{R}_{\mathrm{TE}}(t)\left
  ( \frac{n^2}{n^2+t^2}-\frac{1}{1+t^2}\right ) \mathrm{d}t\nonumber\\ 
&+&\frac{2n}{\sqrt{n^2+1}} \ ,
\end{eqnarray}
and $\tilde{R}_{\mathrm{TE}}$ is the Fresnel reflection coefficient for the
scattering on a straight dielectric interface at imaginary momentum 
\begin{equation}
\tilde{R}_{\mathrm{TE}}(t)=\frac{\sqrt{n^2+t^2}-n^2\sqrt{1+t^2}}{\sqrt{n^2+t^2}+n^2\sqrt{1+t^2}}\
.
\end{equation}   
The plan of the paper is the following. In Sec.~\ref{general} the main
equations
describing the TE modes are reminded. In Sec.~\ref{circle} 
the circular cavity is briefly reviewed: an exact quantization condition is
derived, which allows a direct semiclassical treatment. In
Sec.~\ref{sectweyl}
the first two Weyl terms for the resonance counting function are derived. It
is important to notice that, for TE modes, one can have total transmission of
a ray when the incidence angle is equal to Brewster's angle. This leads to a
special set of 
resonances, which are counted separately in
Sec.~\ref{additional}. Section~\ref{oscillating} is devoted to a brief
derivation
of the oscillating part of the resonance density.  In Sec.~\ref{numerics}
our obtained
formulae are shown to agree well with numerical computation for cavities of
different shapes. In
Appendix~\ref{krein}  another method of deriving the Weyl series for TE
polarization based on Krein's spectral shift formula is presented.  

\section{Generalities}\label{general}

To describe a dielectric cavity correctly one should solve the
$3$-dimensional Maxwell equations. In many applications the transverse
height of 
a cavity, say along the $z$ axis, is much smaller than any other cavity
dimensions.  In such situation the $3$-dimensional problem in a reasonable
approximation can  be reduced to two 2d scalar problems (for each
polarization
of the field) following the so-called effective index approximation, see
e.g. \cite{melanie2007,melanie2008} for more details.   

In the simplest setting, when one ignores  the dependence of the effective
index on frequency,  such 2d approximation consists in using the Maxwell
equations  for an infinite cylinder. It is well known \cite{jackson} that in
this geometry the Maxwell equations are reduced to two scalar Helmholtz
equations inside and outside the cavity 
\begin{eqnarray}
(\Delta +n^2 k^2)\Psi(\vec{x}\,)=0,&\quad& \vec{x} \in {\cal D}\ ,\nonumber\\
(\Delta + k^2)\Psi(\vec{x}\,)=0,&\quad& \vec{x} \notin {\cal D}\ ,
\label{equations}
\end{eqnarray}
where $n$ is the refractive index of the cavity, ${\cal D}$  indicates the
interior of the dielectric cavity, and $\Psi=E_z$ for the TM polarization
and  $\Psi=B_z$ for the TE polarization. 

Helmholtz equations (\ref{equations}) have to be completed by the boundary
conditions.   The field,
$\Psi(\vec{x}\,)$, is continuous across the cavity boundary and its
normal derivatives along both sides of the boundary are related for two
polarizations as below \cite{jackson}
\begin{equation}
\frac{\partial \Psi}{\partial \nu}|_{\mathrm{from\; inside}}=\left \{
\begin{array}{cc} 
\dfrac{\partial \Psi}{\partial \nu}|_{\mathrm{from\; outside}}, &
\mathrm{for\; TM}\ ,\\ \\
n^2 \dfrac{\partial \Psi}{\partial \nu}|_{\mathrm{from\; outside}},
&\mathrm{for\; TE}  \ .
 \end{array}\right . 
 \label{deriv}
\end{equation}
Open cavities have no true discrete spectrum. Instead, we are interested in
the
discrete resonance spectrum, which is defined as the (complex) poles of the
$S$-matrix
for the scattering on a cavity (see e.g. \cite{s_matrix}). It is well known
that the positions of the resonances can be determined directly by the
solution of the problem \eqref{equations} and \eqref{deriv} by imposing  the
outgoing boundary conditions at infinity
\begin{equation}
  \Psi(\vec{x})\propto e^{\ic k |\vec{x}|}\quad  |\vec{x}|\to \infty \ .
\label{infinity}
\end{equation}
The set (\ref{equations})-(\ref{infinity}) admit complex eigen-values  $k$
with Im$\ k<0$, which are the resonances of the
dielectric cavity and are the main object of this paper. Our goal is
to count such resonances for the TE polarization  in the semiclassical
regime. This will
provide us with the analogue of Weyl's law derived for closed systems, see
e.g. \cite{baltes}.

\section{Circular cavity}
\label{circle}

The circular dielectric cavity is the only finite 2d cavity, which permits
an analytical solution. Let  $R$ be the radius
of such a cavity. Writing
$\Psi(\vec{x}\,)=AJ_m(nkr)\mathrm{e}^{\mathrm{i}m\phi}$ inside the
cavity and
$\Psi(\vec{x}\,)=BH_m^{(1)}(kr)\mathrm{e}^{\mathrm{i}m\phi}$ 
outside the cavity, it is plain to check, that in order to fulfill the
boundary conditions, it is necessary that $k$ is determined from the
equation $s_m(x)=0$ with $x=kR$ and 
\begin{equation}
s_m(x)=x\Big
[\frac{1}{n}J_m^{\prime}(nx)H_m^{(1)}(x)-J_m(nx)H_m^{(1)\,\prime}(x)\Big ]
\label{s_m}
\end{equation} 
where $J_m(x)$ (resp. $H_m^{(1)}(x)$) denotes the Bessel function (resp. the Hankel
function of the first kind). Here and below the prime indicates the
derivative
with respect to the argument. Factor $x$ in \eqref{s_m} is introduced for
further convenience.  

Using $J_m(x)=(H_m^{(1)}(x)+H_m^{(2)})/2$ the equation $s_m(x)=0$ can be
rewritten in the form
\begin{equation}
R_m(x)E_m(x)=1\ ,
\label{quantization}
\end{equation}
where
\begin{equation}
E_m(x)=\frac{H_m^{(1)}}{H_m^{(2)}}(nx)
\label{E_m}
\end{equation}
and
\begin{equation}
R_m(x)=\dfrac{
\frac{H_m^{(1)\,\prime}}{H_m^{(1)}}(nx)-n\frac{H_m^{(1)\,\prime}}{H_m^{(1)}}(x)}
{
-\frac{H_m^{(2)\,\prime}}{H_m^{(2)}}(nx)+n\frac{H_m^{(1)\,\prime}}{H_m^{(1)}}(x)}\
.
\label{exactRcircle}
\end{equation}
In the semiclassical limit, $x\to\infty$,  the asymptotic
formula for the Hankel function \cite{bateman} ($0 \le m \le x$) gives
\begin{equation}
H_m^{(1)}(x)\simeq
\frac{\sqrt{2/\pi}}{(x^2-m^2)^{1/4}}\mathrm{e}^{\mathrm{i}\Phi_m(x)}\left
[1+\mathcal{O}(x^{-1})\right ]
\label{hankel}
\end{equation}
where 
\begin{equation}
\Phi_m(x)=\sqrt{x^2-m^2}-m\arccos\left (\frac{m}{x}\right )-\frac{\pi}{4}\ . 
\label{Phi}
\end{equation}
In this way one obtains
\begin{equation}
E_m\underset{x\to\infty}{\longrightarrow}\mathrm{e}^{2\mathrm{i}\Phi_m(nx)}
\end{equation}
and 
\begin{equation}
R_m(x)\underset{x\to\infty}{\longrightarrow} R_{\mathrm{TE}}\left
  (\frac{m}{x} \right )
\label{Rscl}
\end{equation}
where $R_{\mathrm{TE}}$ is the standard TE Fresnel coefficient for the
scattering on an infinite dielectric interface
\begin{equation}
R_{\mathrm{TE}}(t)=\frac{\sqrt{n^2-t^2}-n^2\sqrt{1-t^2}}{\sqrt{n^2-t^2}+n^2\sqrt{1-t^2}}\
.
\label{TE}
\end{equation}
The above formulas mean that in the semiclassical limit,
Eq.~\eqref{quantization} takes the form
\begin{equation}
R_{\mathrm{TE}}\left (\frac{m}{x} \right )
\mathrm{e}^{2\mathrm{i}\Phi_m(nx)}=1
\label{semiclassical}
\end{equation} 
or
\begin{equation}
\Phi_m(n x)=\pi p+\frac{\mathrm{i}}{2}\ln R_{\mathrm{TE}}\left (\frac{m}{x}
\right )
\label{semiclassical_2}
\end{equation}
with integer $p=0,1,2,\ldots$.

In fact, this equation is valid in the semiclassical limit for closed
and open circular cavities with other boundary conditions as well. The only
difference is that, instead of the Fresnel reflection coefficient
$R_{\mathrm{TE}}$, it is necessary to use the reflection coefficient
for the problem under consideration. For example, for closed billiards,
$n=1$ and for Neumann (resp. Dirichlet) boundary conditions $R_m(x)$
in (\ref{Rscl}) equals to $1$ (resp. $-1$). For open dielectric circular
cavity with the TM polarization
$R_m(x)\to R_{\mathrm{TM}}\left (m/x \right )$, where  $R_{\mathrm{TM}}$ is
the usual  Fresnel reflection coefficient for the TM modes \cite{jackson}
\begin{equation}
R_{\mathrm{TM}}(t)=\frac{\sqrt{n^2-t^2}-\sqrt{1-t^2}}{\sqrt{n^2-t^2}+\sqrt{1-t^2}}\
.
\label{TM}
\end{equation}


\section{Weyl terms}\label{sectweyl}
 
Semiclassical formulas like  Eq.~\eqref{semiclassical} are convenient to
obtain the average number of eigenvalues and resonances for closed and
open systems with different boundary conditions.  

Let us consider first the simplest case of a closed billiard with
Neumann boundary conditions for which $R_m(x)=1$. In the semiclassical
regime the eigenvalues for this model are determined from
Eq.~\eqref{semiclassical_2} which reads 
\begin{equation}
\Phi_m(x)=\pi p\ ,
\label{Neumann}
\end{equation}
where $\Phi_m(x)$ is defined in \eqref{Phi} and  $p=0, 1, \ldots $ is an
integer. Therefore, for fixed $m$, the number of eigenvalues less than
$x$ is $ [N_m(x)]$ where $[\;.\; ]$ stands for the integer part and 
\begin{equation}
N_m(x)=\frac{1}{\pi}\Phi_m(x)+1\ .
\end{equation}
 $1$ is added as the integer $p$ in \eqref{Neumann} starts with $0$ but
$[N_m(x)]$ has to begin with $1$. 

Summing over all $m$ leads to the total number of eigenvalues less
than $x$, usually called the counting function. This sum is finite as
the asymptotics \eqref{hankel} is valid when $|m|\leq x$. Finally 
\begin{equation}
N(x)=\sum_{m=-|x|}^{[x]} \Big [N_m(x)\Big ]\ .
\end{equation}
The averaged number of levels is determined from the equation
\begin{equation}
\bar{N}(x)=\sum_{m=-x}^{x}\left ( N_m(x)-\tfrac{1}{2}\right )\ .
\label{averaged}
\end{equation}
With a needed precision one can substitute the summation over $m$ by an
integral and, consequently, the averaged number of eigenvalues for a circular
billiard with Neumann boundary conditions can be approximated as follows 
\begin{equation}
\bar{N}(x)=2\int_{0}^{x}\Big( \frac{1}{\pi
  }(\sqrt{x^2-m^2}-m\arccos\left (\frac{m}{x}\right ))+\frac{1}{4}
\Big )\mathrm{d}m\ . 
\label{neumann_N}
\end{equation}
Using the formula
\begin{equation}
\int_0^1(\sqrt{1-t^2}-t\arccos(t))\mathrm{d}t=\frac{\pi}{8}
\end{equation}
one gets 
\begin{equation}
\bar{N}(x)=\frac{1}{4}x^2+\frac{1}{2}x+\mathcal{O}(1)\ .
\end{equation}
As for the circle the area is $\mathcal{A}=\pi R^2$ and the perimeter is
$\mathcal{L}=2\pi R$,
these results can be rewritten in the standard form \cite{baltes}
\begin{equation}
\bar{N}(k)=\frac{\mathcal{A}\, k^2 }{4\pi}+r \frac{\mathcal{L}\,
k}{4\pi}+\mathcal{O}(1)\ 
\label{weyl}
\end{equation} 
with $r=1$. For Dirichlet boundary conditions similar arguments show that
$1/4$ in \eqref{neumann_N} is substituted by $-1/4$ and  $r=-1$, as it should
be \cite{baltes}. 

\begin{figure*}
 \includegraphics[angle=-90, width=.9\linewidth]{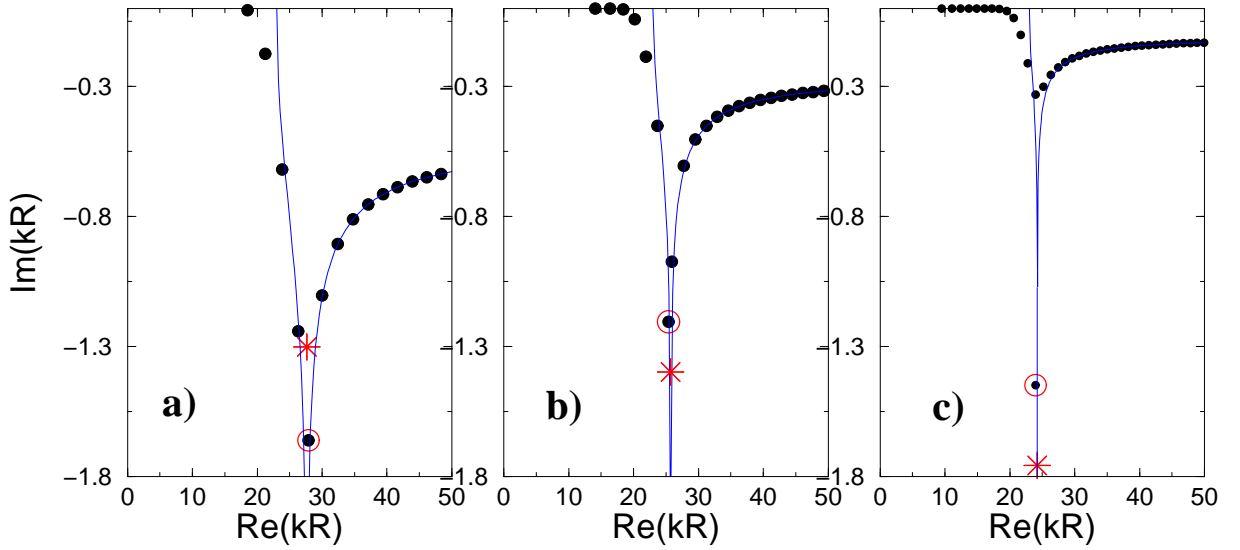}
 \caption{(Color online). The black circles are the exact positions of the
resonances  with $m=23$
   and $n=1.5$ (a), $n=2$ (b), and $n=3$ (c). The blue lines indicate the
   approximation \eqref{imaginary_approx}. The additional levels are
   encircled by the red circles. The red stars show the approximate formula
   \eqref{tilde_x}. } 
 \label{m_23_app}
 \end{figure*}

For open cavities Eq.~(\ref{semiclassical}) gives complex solutions
(resonances)
$k=k_1+\mathrm{i}k_2$ with negative imaginary part, $k_2<0$. In the
semiclassical
limit for all investigated cases one has $|k_2|\ll k_1$. Separating the
imaginary
and real parts in \eqref{semiclassical_2} and using that  
\begin{equation}
\frac{\partial }{\partial x}\Phi_m(x)=\sqrt{1-\frac{m^2}{x^2}}
\end{equation}
one gets that in the first order in $k_2$ the real part of the
resonance position, $k_1$ (or $x_1=k_1R$), is determined from the real
equation
similar to \eqref{semiclassical_2}
\begin{equation}
\Phi_m(n x_1)+\delta(m/x_1)=\pi p,\quad p \textrm{ integer}, 
\label{real_equation}
\end{equation}
where $2\delta (t)$ is the argument of the reflection coefficient
\begin{equation}
R(t)=|R(t)| \mathrm{e}^{2\mathrm{i}\delta(t)}\ .
\end{equation}
In the same approximation the imaginary part of the resonance position,
$k_2$ is
\begin{equation}
k_2 R=\frac{\ln |R(x_1)|}{2\sqrt{n^2-m^2/x_1^2}}\ .
\label{imaginary_approx}
\end{equation}
 This semiclassical  approximation is quite good even for not too large $m$
as indicated in Fig.~\ref{m_23_app}.

The above arguments demonstrate that the total number of resonances can be
calculated from
the real equation~\eqref{real_equation}. As in \eqref{averaged}  one
concludes
that the mean number of resonances with real part $x_1$ less
than $x$ is given by the expression 
\begin{equation}
\bar{N}(x)=\sum_{m=-nx}^{nx}\left ( N_m(x)-\tfrac{1}{2}\right )
\end{equation}
with 
\begin{equation}
N_m(x)=\frac{1}{\pi}\Phi_m(nx)+\frac{1}{\pi}\delta\left (\frac{m}{x}\right
)+1\ .
\end{equation}
Consider first the case of TM modes. The reflection coefficient in this case
is given by \eqref{TM} and one has
\begin{equation}
\delta_{\mathrm{TM}}(t)=\left \{ \begin{array}{rr} -\arctan \left
(\dfrac{\sqrt{t^2-1}}{\sqrt{n^2-t^2}}\right ),& 1\leq t\leq n\\ \\ 0,& 0\leq
t\leq 1\end{array}\right . \ .
\end{equation}
Therefore
\begin{eqnarray}
\bar{N}(x)&=&2\int_{0}^{nx} \frac{1}{\pi }\left [\sqrt{n^2
x^2-m^2}-m\arccos\left (\frac{m}{nx}\right ) \right ]\mathrm{d}m\nonumber\\
&+&2\int_{0}^{nx}\left  [ \frac{1}{\pi } \delta_{\mathrm{TM}}\left
(\frac{m}{x}\right )+\frac{1}{4} \right ]\mathrm{d}m
= \frac{n^2}{4}x^2+\frac{n}{2}x\nonumber\\
&-&\frac{2x}{\pi}\int_{1}^n\arctan \left
(\frac{\sqrt{t^2-1}}{\sqrt{n^2-t^2}}\right )\mathrm{d}t\ .
\end{eqnarray}
By integration by parts and contour deformation it is easy to check that 
\begin{eqnarray}
& &2\int_{1}^n\arctan \left (\frac{\sqrt{t^2-1}}{\sqrt{n^2-t^2}}\right
)\mathrm{d}t=\frac{\pi}{2}(n-1)\nonumber\\
&-&\int_0^{\infty}\tilde{R}_{\mathrm{TM}}(t)\left (
\frac{n^2}{n^2+t^2}-\frac{1}{1+t^2}\right ) \mathrm{d}t
\label{integrals_TM}
\end{eqnarray}
where $\tilde{R}_{\mathrm{TM}}(t) $ is the same as \eqref{TM} but for pure
imaginary argument
\begin{equation}
\tilde{R}_{\mathrm{TM}}(t)\equiv {R}_{\mathrm{TM}}(\ic t)=
\frac{\sqrt{n^2+t^2}-\sqrt{1+t^2}}{\sqrt{n^2+t^2}+\sqrt{1+t^2}}\ .
\label{TM_imaginary}
\end{equation}
Finally, these considerations lead to the expression similar to \eqref{weyl}
\begin{equation}
\bar{N}_{\mathrm{TM}}(k)=\frac{\mathcal{A}\, n^2 k^2
}{4\pi}+r_{TM}\frac{\mathcal{L}\, k}{4\pi}+o(k) 
\label{weyl_TM}
\end{equation}
where
\begin{equation}
r_{TM}(n)=1+\frac{1}{\pi}\int_{-\infty}^{\infty}\tilde{R}_{\mathrm{TM}}(t)\left
  ( \frac{n^2}{n^2+t^2}-\frac{1}{1+t^2}\right ) \mathrm{d}t 
\label{r_TM}
\end{equation}
which agrees with the result in \cite{trace1} obtained by a different
method.  

Consider now  TE modes. In this case  the reflection coefficient is given by
\eqref{TE} and its argument is
\begin{equation}
\delta_{\mathrm{TE}}(t)=\left \{ \begin{array}{rr} -\arctan \left
      (\dfrac{n^2\sqrt{t^2-1}}{\sqrt{n^2-t^2}}\right ),& 1\leq t\leq n\\ \\0,&
    t^*\leq t\leq 1\\  \\
-\dfrac{\pi}{2},& 0\leq t\leq t^*\end{array}\right . 
\label{phase_delta}
\end{equation}
where $t^*$ corresponds to the zero of the TE reflection coefficient
(Brewster's angle), $R_{\mathrm{TE}}(t^*)=0$,
\begin{equation}
t^*=\frac{n}{\sqrt{n^2+1}}\ .
\label{t_star}
\end{equation}
Using these values we get
\begin{equation}
\bar{N}_{\mathrm{TE}}(k)=\frac{\mathcal{A}\, n^2 k^2
}{4\pi}+r_{TE}\frac{\mathcal{L}\, k}{4\pi}+o(k)\ 
\label{weyl_TE}
\end{equation}
where $r_{TE}$ is given by the following expression
\begin{eqnarray}
r_{TE}(n)&=&-\frac{4}{\pi}\int_{1}^n \arctan \left
(\frac{n^2\sqrt{t^2-1}}{\sqrt{n^2-t^2}}\right )\mathrm{d}t\nonumber\\
&+&n-2\int_0^{t^*}\mathrm{d}t \ .
\end{eqnarray}
Similar to \eqref{integrals_TM} one can prove that
\begin{eqnarray}
& &2\int_{1}^n\arctan \left (n^2 \frac{\sqrt{t^2-1}}{\sqrt{n^2-t^2}}\right
)\mathrm{d}t=\frac{\pi}{2}(n-1)\nonumber\\
&-&\int_0^{\infty}\tilde{R}_{\mathrm{TE}}(t)\left (
\frac{n^2}{n^2+t^2}-\frac{1}{1+t^2}\right ) \mathrm{d}t
\label{integrals_TE}
\end{eqnarray}
where, as above, $\tilde{R}_{\mathrm{TE}}(t) $  is the TE reflection
coefficient \eqref{TE} analytically continued to imaginary $t$ 
\begin{equation}
\tilde{R}_{\mathrm{TE}}(t)\equiv R_{\mathrm{TE}}(\ic t)=
\frac{\sqrt{n^2+t^2}-n^2\sqrt{1+t^2}}{\sqrt{n^2+t^2}+n^2\sqrt{1+t^2}}\ . 
\label{TE_imaginary}
\end{equation}
Combining all terms together we obtain that 
\begin{eqnarray}
r_{TE}(n)&=&1+\frac{1}{\pi}\int_{-\infty}^{\infty}\tilde{R}_{\mathrm{TE}}(t)\left
( \frac{n^2}{n^2+t^2}-\frac{1}{1+t^2}\right ) \mathrm{d}t\nonumber\\
&-&\frac{2n}{\sqrt{n^2+1}} \ .
\label{r_TE}
\end{eqnarray}
The first two terms are the same as for TM modes \eqref{r_TM} but with TE
reflection coefficient. The last term is the new one related to
the change of the sign of the TE reflection coefficient.  

Higher order terms in Weyl's expansions \eqref{weyl_TM} and \eqref{weyl_TE}
are not yet calculated so  we prefer to use a conservative  estimate of them
as $o(k)$ though all  numerical checks  suggest that for smooth boundary
cavities it is $\mathcal{O}(1)$.  
\section{Additional resonances}\label{additional}

Formula \eqref{r_TE} is the correct description for the resonances
whose real part of the eigen-momentum $k$ corresponds to non-zero reflection
coefficient (i.e. $m/x_1\neq t^*$). This is due to the fact that when the
reflection coefficient is zero  its phase is not defined. 
For TE modes there is a special branch of resonances for which
semiclassically
the real part does obey
$m/x_1=t^*$. The existence of such additional  resonances were first discussed in
a different context in \cite{sieber}. 

The approximate positions of these resonances can be calculated as follows.
Assume that the resonances have a large imaginary
part. As $H_m^{(2)}(x-\mathrm{i}\tau)$ tends to zero when
$\tau\to\infty$ one can approximate Eq.~\eqref{s_m} by 
\begin{equation}
\tilde{s}_m(x)=0,\quad\tilde{s}_m(x)=\frac{H_m^{(1)\, \prime
}}{H_m^{(1)}}(nx)-n\frac{H_m^{(1)\, \prime }}{H_m^{(1)}}(x)\ .
\end{equation}
From the asymptotic \eqref{hankel} it follows that
\begin{equation}
\frac{H_m^{(1)\, \prime
  }}{H_m^{(1)}}(x)\underset{x\to\infty}{\longrightarrow}
\mathrm{i}\sqrt{1-\frac{m^2}{x^2}}-\frac{x}{2(x^2-m^2)}\ . 
\end{equation} 
Using this expression one concludes that the solution of the equation
$\tilde{s}_m(\tilde{x}_m)=0$ has the form
\begin{equation}
\tilde{x}_m\approx
\frac{\sqrt{n^2+1}}{n}m-\mathrm{i}\frac{(n^2+1)^{3/2}}{2n^2}\ .
\label{tilde_x}
\end{equation}
This approximation is better for large $n$ when the imaginary part is
large but it gives reasonable results even for $n$ of the order of
$1$. In practice one may use \eqref{tilde_x}  as the initial value for any
root search algorithm (cf. Fig.~\ref{m_23_app}).  

From  \eqref{tilde_x}  it follows that the ratio $m/\tilde{x}_m$ tends
to $t^*$ defined in \eqref{t_star} so these resonances are not taken
explicitly into account in Eq.~(\ref{r_TE}).  Their number can be
estimated as follows. The discussed resonances correspond to waves
propagating
along the boundary whose direction  forms an angle with the normal exactly
equal to
Brewster's angle 
\begin{equation}
\sin \theta_{\mathrm{B}}=\frac{n}{\sqrt{n^2+1}}
\end{equation}
If the length of the boundary is $\mathcal{L}$, the possible values for the
momenta of such states in the semiclassical limit are
\begin{equation}
k_m \sin \theta_{\mathrm{B}}=\frac{2\pi}{\mathcal{L}}m
\label{kbrewster}
\end{equation}
with integer $m=0,\pm 1,\pm 2,\ldots$. Therefore, the number of
additional resonances related with Brewster's angle is 
\begin{equation}
N_{\mathrm{add}}(k)\approx \frac{\mathcal{L}\, k}{\pi}\frac{n}{\sqrt{n^2+1}}\ .
\end{equation}
Comparing it with Eq.~\eqref{r_TE} we conclude that the second term in
the Weyl expansion for the averaged number of resonances for TE
polarization is the following 
\begin{eqnarray}
r_{TE}&=&1+\frac{1}{\pi}\int_{-\infty}^{\infty}\tilde{R}_{\mathrm{TE}}(t)\left
  ( \frac{n^2}{n^2+t^2}-\frac{1}{1+t^2}\right ) \mathrm{d}t\nonumber\\ 
&\pm &\frac{2n}{\sqrt{n^2+1}} 
\label{r_TE_final}
\end{eqnarray} 
where the plus sign is used when the above additional resonances are
taken into account and the minus sign corresponds to the case when
these resonances are ignored. 

For small values of $n$ the additional
resonances are mixed with other resonances and their separation seems
artificial. For large $n$ the additional branch of resonances is well
separated from the main body of resonances and one can decide not to
take them into account. In such a case, the minus sign has to be used
in \eqref{r_TE_final} (see below Section~\ref{numerics}). 

When the cavity remains invariant under a group of symmetry it is often
convenient to split resonances according to their symmetry
representations. For reflection symmetries it is equivalent  to  consider a
smaller cavity  where along parts of the boundary one has to impose either
Dirichlet or Neumann boundary conditions. In this case the total boundary
contribution to the average counting function $\bar{N}(k)$ is given by the
general formula 
\begin{equation}
\frac{1}{4\pi}\left ( n(L_N-L_D)+r_{TE}(n)L_0 \right )k\ .
\label{reflection_symmetry}
\end{equation} 
Here $L_N$ and $L_D$ are the lengths of the boundary parts with respectively
Neumann and Dirichlet boundary conditions and $L_0$ is the length of the true
dielectric interface. It is this formula, which will be used in
Section~\ref{numerics} for dielectric cavities in the shape of a square and
a stadium. 


\section{Oscillating part of the trace formula}\label{oscillating}

The  quantization conditions \eqref{s_m} or \eqref{quantization}   permit
also to obtain  the resonance trace formula for a circular dielectric
cavity. Let $k_j=k_{1j}-\mathrm{i}k_{2j}$ be resonance eigen-momenta. Define
the density of resonances as follows
\begin{equation}
d(k)=-\frac{1}{\pi}\mathrm{Im}\, \sum_j \frac{1}{k-k_j}=\frac{1}{\pi} \sum_j
\frac{k_{2j}}{(k-k_{1j})^2+k_{2j}^2}\ .
\label{density}
\end{equation} 
In general, if $x_j$ are the zeros of a certain function $F(x)$   which has
no other singularities  then the density of these zeros \eqref{density}
formally is given by the following expression 
\begin{equation}
d(x)=-\frac{1}{\pi}\mathrm{Im}\frac{F^{\prime}(x)}{F(x)}\ .
\end{equation}
In the semiclassical limit $k\to \infty$ it is sufficient to consider the
semiclassical formula \eqref{semiclassical} i.e. $F(x)=\prod_m F_m(x)$ and 
\begin{equation}
F_m(x)=1-R_{\mathrm{TE}}\left (\frac{m}{x} \right )
\mathrm{e}^{2\mathrm{i}\Phi_m(nx)}\ .
\end{equation}
A more careful discussion is performed in Appendix~\ref{krein}. In such a
manner one gets
\begin{eqnarray}
&&d^{(osc)}(k)=\frac{2R}{\pi}\sum_{m=-\infty}^{\infty}\mathrm{Re}\,
\sqrt{n^2-\frac{m^2}{ x^2}}\,
\frac{R_{m}\mathrm{e}^{2\mathrm{i}\Phi_m(nx)}}{1-R_{m}\mathrm{e}^{2\mathrm{i}\Phi_m(nx)}
}\nonumber\\
&&=\frac{R}{\pi} \sum_{m=-\infty}^{\infty}\sqrt{n^2-\frac{m^2}{ x^2}} \,
\sum_{r=1}^{\infty}R_{m}^r\mathrm{e}^{2r\mathrm{i}\Phi_m(nx)}+\mathrm{c.c.}
\label{d_osc}
\end{eqnarray}
Here $R_{m}=R_{\mathrm{TE}}(m/x)$ is the Fresnel reflection coefficient for
TE polarization \eqref{TE}. 

The further steps are as usual, see e.g. \cite{trace1}. Using the Poisson
summation formula
\begin{equation}
\sum_{m=-\infty}^{\infty}f(m)=\sum_{M=-\infty}^{\infty}
\int_{m=-\infty}^{\infty}\mathrm{e}^{2\pi \mathrm{i}Mm} f(m)\mathrm{d}m
\end{equation}
the expression \eqref{d_osc} becomes
\begin{eqnarray}
&&d^{(osc)}(k)=\\
&&\frac{R}{\pi}
\sum_{M=-\infty}^{\infty}\int_{m=-\infty}^{\infty}\mathrm{d}m 
\sqrt{n^2-\frac{m^2}{ x^2}} \,
\sum_{r=1}^{\infty}R_{m}^r\mathrm{e}^{\mathrm{i}S_{M,r}(m)}+\mathrm{c.c.}
\nonumber
\end{eqnarray} 
where the action is
\begin{equation}
S_{M,r}(m)=2\pi Mm +2r\Phi_m(nx)\ .
\end{equation}
When $k\to\infty$ the dominant contribution to the integral is due to saddle
point solutions $m_{sp}$ determined from the equation $\partial
S_{M,r}(m)/\partial m=0$. It is plain that 
\begin{equation}
m_{sp}=nx\cos(\theta_{M,r})
\label{saddle}
\end{equation}
with $ \theta_{M,r}=\pi M/r$. This saddle point corresponds geometrically to
a periodic orbit of the circle in the shape of regular polygon with $r$
vertices going around the center $M<r$ times. Expanding the action
$S_{M,r}(m)$ around the saddle point \eqref{saddle} one gets
\begin{equation}
S_{M,r}(m_{sp}+\delta m)\approx nkl_p-\frac{\pi}{2}r +\frac{r}{nx\sin
\theta_{M,r}}(\delta m)^2 \ .
\end{equation}
Here $l_p=2rR\sin \theta_{M,r}$ is the classical length of the periodic
orbit determined by $M$ and $r$. 

In the end one gets the trace formula for the resonances of the circular
dielectric cavity in the  form
\begin{equation}
d^{(osc)}(k)=2k\frac{n^{3/2}}{\pi}\sum_{M,r}\frac{\mathcal{A}_p}{\sqrt{2\pi
k l_p}}R_p^r\mathrm{e}^{\mathrm{i}[nkl_p-r\pi/2+\pi/4]} +\mathrm{c.c.}
\end{equation} 
where $\mathcal{A}_p=\pi R^2\sin^2\theta_{M,r}$ is the area occupied by a
given periodic orbit family, $R_p=R_{\mathrm{TE}}(n\cos \theta_{M,r})$
is the
Fresnel reflection coefficient for the TE scattering with an angle equal to
the
reflection angle, $\theta_{M,r}$, for the given periodic orbit.  

Repeating the arguments presented in \cite{trace1} we argue that in general
the oscillating part of the resonance trace formula in the strong
semiclassical limit has the form of the sum over all classical periodic
orbits 
\begin{equation}
d(E)=\sum_p d_p(E)+\textrm{ c.c. }
\end{equation}
where contribution of an individual orbit depends on the orbit considered
\begin{itemize}
\item For an isolated primitive periodic orbit $p$ repeated $r$ times
  \begin{equation}
    \label{dE_chao}
    d_p(E)=\frac{n l_p}{\pi \ |\det(M_p^r-1)|^{1/2}} R_p^r \mathrm{e}^{\ic r
n k l_p-\ic r \mu_p\pi/2}
  \end{equation}
where  $l_p,M_p, \mu_p$, $R_p$  are, respectively, the length, the monodromy
matrix, the Maslov index, and  the total TE Fresnel reflection coefficient
for the chosen primitive periodic orbit.
\item For a primitive periodic orbit  family  
  \begin{equation}
    \label{dE_int}
  d_p(E)=\frac{n^{3/2}\mathcal{A}_p}{\pi \ \sqrt{2\pi k l_p}} \langle R_p^r
\rangle \mathrm{e}^{\ic n k l_p-\ic r \mu_p\pi/2}
  \end{equation}
where   $\mathcal{A}_p$ is the area covered by one periodic orbit family,
$\langle R_p^r \rangle $ is the mean value of the TE Fresnel reflection
coefficient averaged over a periodic orbit family. 
\end{itemize}
The only difference with corresponding results derived  in \cite{trace1}  is
that the TE reflection coefficient is used instead of the TM coefficient.

\section{Numerical verification}\label{numerics}

\begin{figure*}[t]
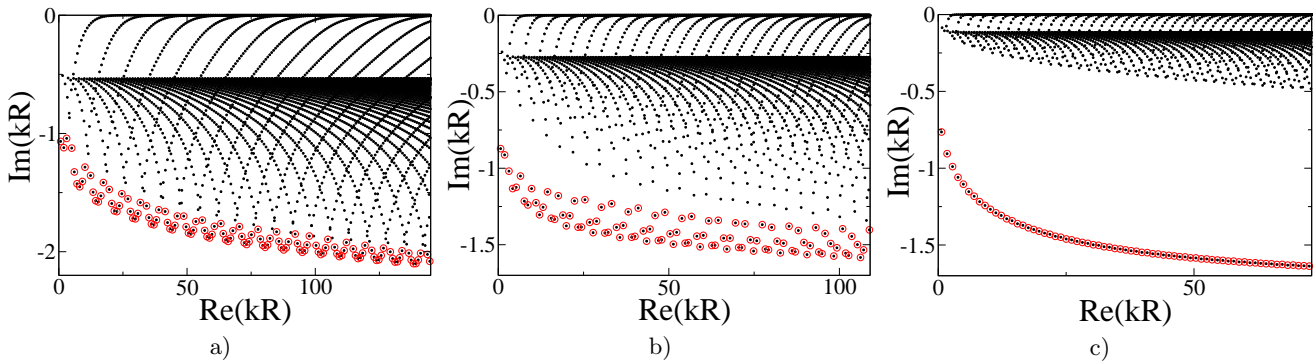

\begin{minipage}{.32\linewidth}
\includegraphics[angle=0, width=.99\linewidth,clip]{fig2a.eps}
\centering a)
\end{minipage}
\begin{minipage}{.32\linewidth}
\includegraphics[angle=0, width=.99\linewidth,clip]{fig2b.eps}
\centering b)
\end{minipage}
\begin{minipage}{.32\linewidth}
\includegraphics[angle=0, width=.99\linewidth,clip]{fig2c.eps}
\centering c)
\end{minipage}
\caption{(Color online). Resonance spectra for TE modes of the circular
dielectric cavity with
  $n=1.5$ (a), $n=2$ (b), and $n=3$ (c). The red circles encircle  
   the additional branch of resonances obtained by choosing initial
conditions  \eqref{tilde_x} and running a root searching routine to solve
the equation $s_m(x)=0$ with $s_m(x)$ given by \eqref{s_m}. The range along
the $x$ axis are chosen such that every plot contains around
12000 resonances (counted with multiplicity).} 
\label{circle_n_1.5_2_3}
\end{figure*}

The numerical calculations of the resonance spectrum for the TE
modes of the circular dielectric cavity is presented in
Fig.~\ref{circle_n_1.5_2_3}.  Notice that when the cavity
refraction index $n$ increases the additional branch of the resonances
(\ref{tilde_x}) separates more and
more from the main part of the spectrum.  

\begin{figure*}[b]
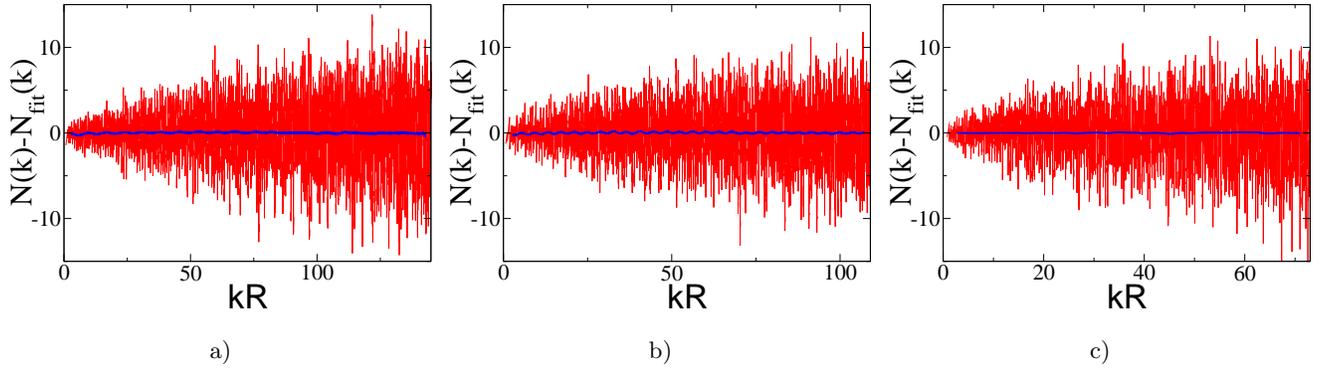

\begin{minipage}{.32\linewidth}
\includegraphics[ width=.98\linewidth,clip]{fig3a.eps}
\begin{center} a)\end{center}
\end{minipage}
\begin{minipage}{.32\linewidth}
\includegraphics[ width=.98\linewidth,clip]{fig3b.eps}
\begin{center} b)\end{center}
\end{minipage}
\begin{minipage}{.32\linewidth}
\includegraphics[ width=.98\linewidth,clip]{fig3c.eps}
\begin{center} c)\end{center}
\end{minipage}
\caption{(Color online) Difference between the exact number of resonances and
  the fit (\ref{Nfitcirc}) for 
  $n=1.5$ (a), $n=2$ (b), and $n=3$ (c). In the latter case the
  additional resonances  in
  Fig.~\ref{circle_n_1.5_2_3} c) are not taken into account. The blue solid
thick
  line indicates the difference averaged over a large interval.} 
\label{circle_fit_n_1.5_2_3}
\end{figure*}

In Fig.~\ref{circle_fit_n_1.5_2_3} we plot the difference between the
function counting the numerically computed resonances 
resonances 
with a real part less than $k$ (with
radius $R=1$) and the best fit to it of the form, see \eqref{weyl_TE},  
\begin{equation}
N_{\mathrm{fit}}(k)=\frac{n^2}{4} (kR)^2+a_1 (kR)+a_0  \label{Nfitcirc}
\end{equation}
where $a_1$ and $a_0$ are fitting parameters. 
  
For $n=1.5$ and $n=2$ we consider all resonances including the additional
branch. For $n=3$ this branch is quite far from the other resonances
(cf. Fig.~\ref{circle_n_1.5_2_3} c)) and it is natural  not to include
it in the counting. The fitted values of the parameters for these three
cases are the following 
\begin{equation}
\begin{array}{lll }n=1.5,& a_1=1.246,& a_0=-0.66\\n=2,& a_1=1.122, &
  a_0=-0.50\\ n=3,& a_1=-1.189, & a_0=0.12\end{array} . 
\end{equation}
The term $a_1$ has to be compared with the theoretical prediction which
follows from \eqref{r_TE_final} (used with plus sign for $n=1.5$ and
$n=2$, and with minus sign for $n=3$)  
\begin{equation}
\begin{array}{ll}n=1.5,& a_1^{\mathrm{th}}=1.247\\n=2,&
  a_1^{\mathrm{th}}=1.124 \\ n=3,& a_1^{\mathrm{th}}=-1.190 \end{array} . 
\end{equation}  
The agreement with our numerical calculations is very good. 

\begin{figure*}
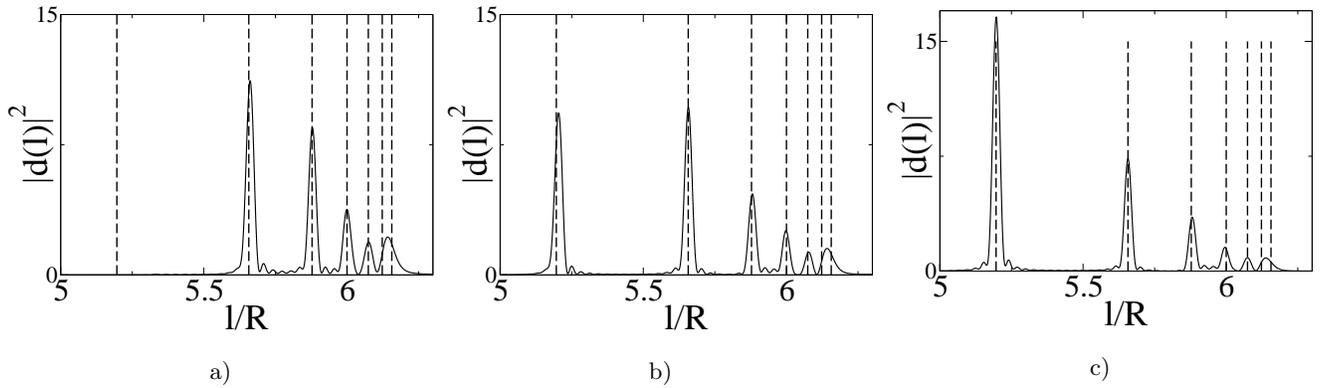

\begin{minipage}{.32\linewidth}
\includegraphics[ width=.99\linewidth,clip]{fig4a.eps}
\begin{center} a)\end{center}
\end{minipage}
\begin{minipage}{.32\linewidth}
\includegraphics[ width=.99\linewidth,clip]{fig4b.eps}
\begin{center} b)\end{center}
\end{minipage}
\begin{minipage}{.32\linewidth}
\includegraphics[ width=.99\linewidth,clip]{fig4c.eps}
\begin{center} c)\end{center}
\end{minipage}
\caption{Density of periodic orbit length (Fourier transform of
  the resonance density) for 
  $n=1.5$ (a), $n=2$ (b), and $n=3$ (c). The vertical lines stand for the
  length of the shortest periodic orbits of the circular cavity, from left to right: triangle,
square, pentagon, hexagon, heptagon and octagon.} 
\label{d_l_circle}
\end{figure*}

In Fig.~\ref{d_l_circle} the Fourier transform of the resonance density for
the circular dielectric cavity with different values of the refractive index
is displayed.  As expected from the trace formula, this quantity has peaks
at the length of classical periodic orbits of the
circle. Notice especially that the triangular orbit is not confined for
$n=1.5$. Hence the Fresnel reflection coefficient is small and induces
damping,
which can be clearly seen in Fig.~\ref{d_l_circle}~a). As the index grows it
is also shown that the contribution of short-period  orbits become closer and
closer to the one of the closed billiard.

\begin{figure*}
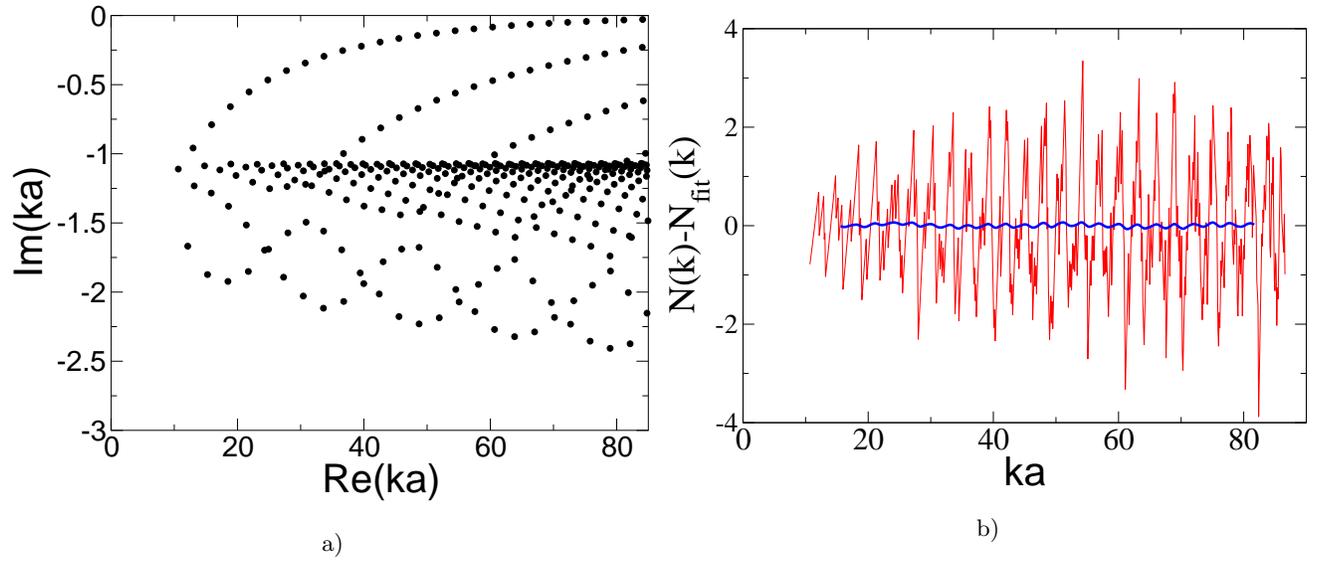

\begin{minipage}{.48\linewidth}
\includegraphics[ width=.99\linewidth,clip]{fig5a.eps}
\begin{center} a)\end{center}
\end{minipage}
\begin{minipage}{.48\linewidth}
\includegraphics[ width=.99\linewidth,clip]{fig5b.eps}
\begin{center} b)\end{center}
\end{minipage}
\caption{(Color online) a) Resonance spectrum for the dielectric square with
$n=1.5$
  for the $(-,-)$ symmetry class. b) The difference between the total number
  of resonances and the best quadratic fit. The blue solid thick
  line indicates the difference averaged over a large interval.} 
\label{square_fig}
\end{figure*}

In Fig.~\ref{square_fig}~a) we present the spectrum of the TE resonances
for the square cavity of side $a=1$ with $(-\,-)$ symmetry along the
diagonals. For such cavity  the fit
function similar to (\ref{Nfitcirc}) is 
\begin{equation}
  \label{Nfitsq}
  N_{\mathrm{fit}}(k)=\frac{n^2}{16\pi} (ka)^2+a_1 (ka)+a_0
\end{equation}
and the best fit gives, see Fig.~\ref{square_fig}~b),   
\begin{equation}
a_1=0.0304,\qquad a_0=-5.22\ .
\end{equation}
The theoretical prediction for this symmetry class is obtained from
\eqref{reflection_symmetry}  $a_1^{\mathrm{th}}=0.0297$ which agrees well
with
the numerical calculations. 

\begin{figure*}
\begin{minipage}{.48\linewidth}
\includegraphics[ width=.99\linewidth,clip]{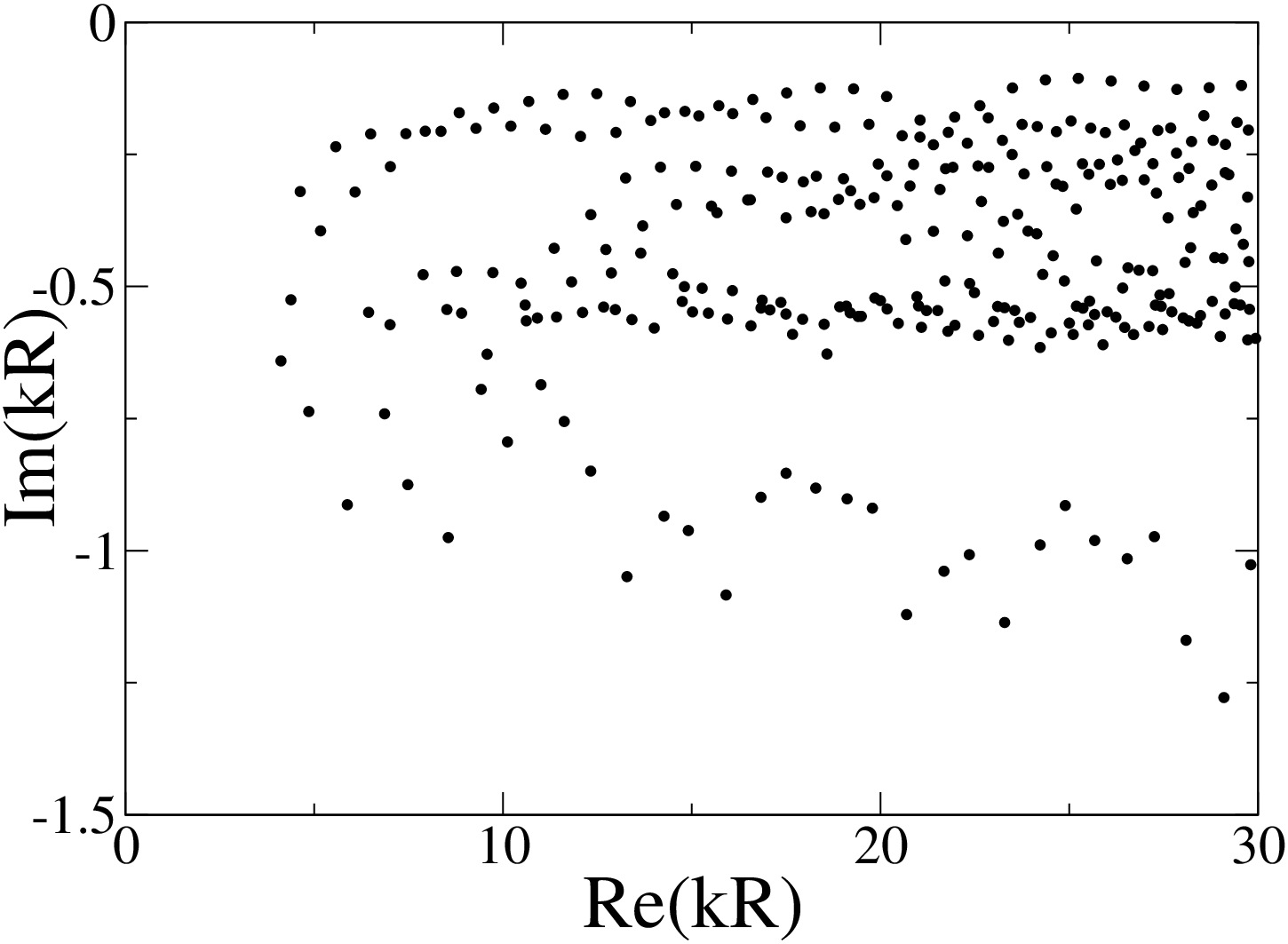}
\begin{center} a)\end{center}
\end{minipage}
\begin{minipage}{.48\linewidth}
\includegraphics[ width=.99\linewidth,clip]{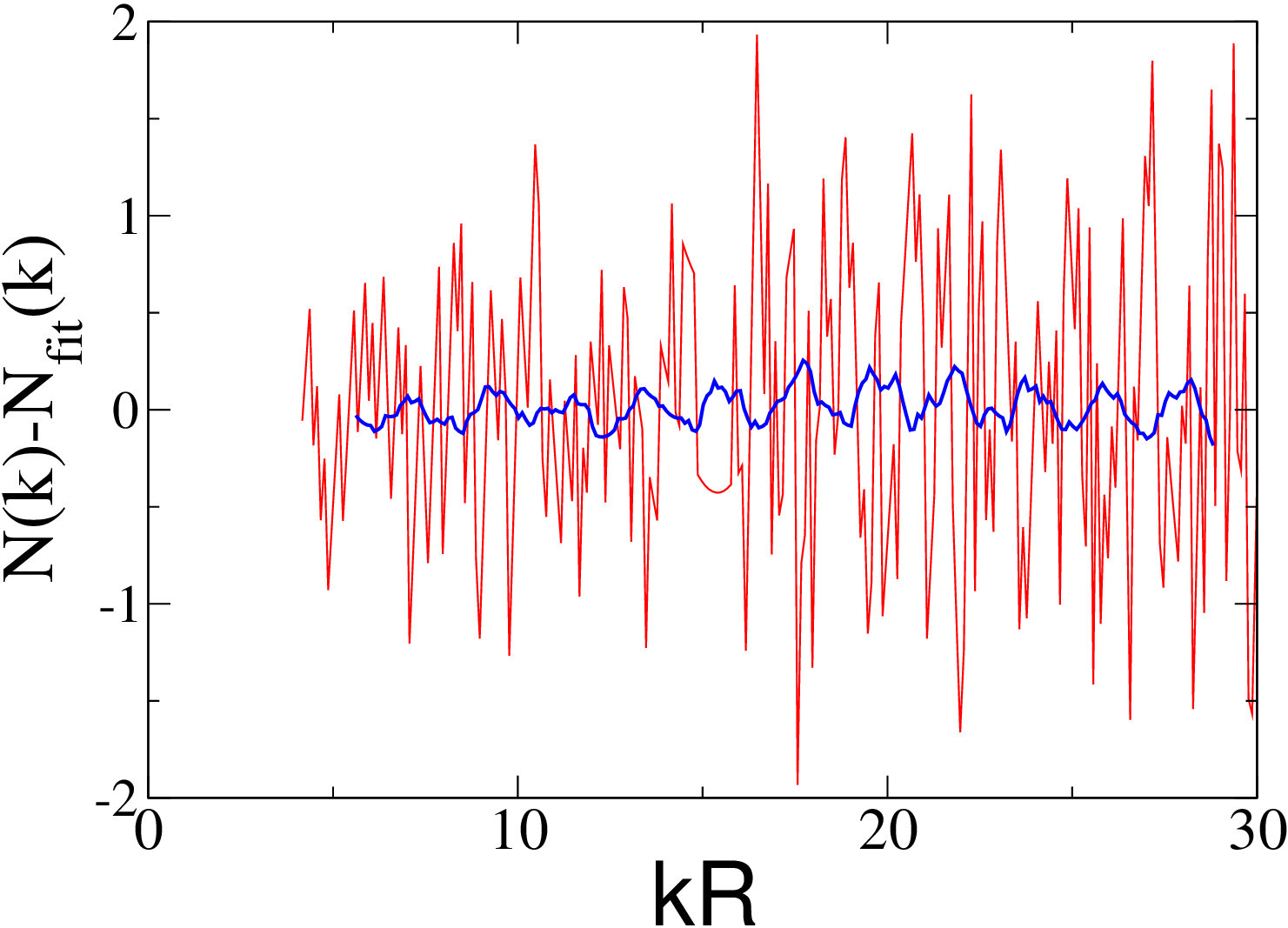}
\begin{center} b)\end{center}
\end{minipage}
\caption{(Color online) a) Resonance spectrum for the dielectric stadium
with $n=1.5$
  for the $(-,-)$ symmetry class. b) The difference between the total number
  of resonances and the best quadratic fit. The blue solid thick
  line indicates the difference averaged over a large interval. } 
\label{stadium_fig}
\end{figure*}

Finally the same procedure was done for the dielectric stadium consisted of
two half-circles of radius $R$ connected by a rectangle with sides $2\alpha R$
and $2R$ where $\alpha$ called the aspect ratio of the stadium. The 
calculations were 
restricted to the symmetry class such that the 
associated function vanishes along both symmetry axis of the
stadium, which is again called $(-,-)$ symmetry class. The resonance
spectrum is presented in Fig.~\ref{stadium_fig}~a). 

The fit function is now 
\begin{equation}
  \label{Nfitst}
   N_{\mathrm{fit}}(k)=\frac{n^2}{4\pi}\left(\alpha+\frac{\pi}{4}\right)
(kR)^2+a_1 (kR)+a_0
\end{equation}
where  the aspect ratio $\alpha$  has been taken to $1$
in the numerical calculations. The best fit gives, see
Fig.~\ref{stadium_fig}~b),
\begin{equation}
  a_1=0.150,\quad a_0=-5.24\ ,
\end{equation}
which agrees  well with the prediction for this symmetry class:
$a_1^{\mathrm{th}}=0.152$ (cf. \eqref{reflection_symmetry}).

\section{Summary}

Trace formulas are the main tool of the semiclassical description of
multi-dimensional quantum problems. For closed systems the trace formulas
relate two objects: quantum density of discrete states and a sum
over classical periodic orbits
\begin{eqnarray}
& &d(E)\equiv \sum_n\delta(E-E_n)\nonumber\\
& &\approx \bar{d}(E)+\sum_{\mathrm{period.\
orbits}}A_p\mathrm{e}^{\mathrm{i}S_p(E)/\hbar}+\mathrm{c.c.}
\label{usual_trace}
\end{eqnarray}
where $S_p(E)$ is the classical action over a periodic orbit and $\bar{d}(E)$
is the mean density of eigen-energies, averaged over a small window around
$E$. For 2d billiards with area $\mathcal{A}$ and perimeter $\mathcal{L}$
this averaged density of states is 
\begin{equation}
\bar{d}(E)=\frac{\mathcal{A}}{4\pi}+r\frac{\mathcal{L}}{8\pi
\sqrt{E}}+o(E^{-1/2})
\end{equation}
where $r=1$ for the Neumann boundary conditions and $r=-1$ for the Dirichlet
ones. 

For open quantum models the true eigen-energy spectrum is continuous and the
main object of interest is the discrete spectrum of resonances defined as the
poles of the $S$-matrix in the complex plane: $E_n=e_n-\mathrm{i}\Gamma_n/2$
with $e_n$ and $\Gamma_n$ real. The real part of the resonance energy, $e_n$,
gives the position of the resonance while its imaginary part, $\Gamma_n$,
determines the resonance width.  

The analogue of the trace formula for open systems has the form similar to
\eqref{usual_trace}
\begin{eqnarray}
& &\frac{1}{\pi}\sum_n \dfrac{\Gamma_n/2}{(E-e_n)^2+\Gamma_n^2/4}\nonumber\\
& &\approx \bar{d}(E)+\sum_{\mathrm{period.\
orbits}}A_p\mathrm{e}^{\mathrm{i}S_p(E)/\hbar}+\mathrm{c.c.}\ .
\end{eqnarray}
In Ref.~\cite{trace1}  such type of formula has been obtained for a 2d
dielectric cavity with transverse magnetic polarization of the field. 
Here we derive the trace  formula for a 2d dielectric cavity but with
boundary
conditions corresponding to the transverse electric polarization of the
electromagnetic field. As expected, the oscillating part of this trace
formula
is given by the usual periodic orbits weighted in the leading semiclassical
order by  the Fresnel coefficient corresponding to TE reflection on the
cavity
boundary \eqref{dE_chao}, \eqref{dE_int}.  

Our main result is the expression for the average resonance density of a
dielectric cavity with area $\mathcal{A}$, perimeter $\mathcal{L}$, and
refraction index $n$
\begin{equation}
\bar{d}(E)=\frac{\mathcal{A}n^2}{4\pi}+r_{\mathrm{TE}}(n)\frac{\mathcal{L}}{8\pi
\sqrt{E}}+o(E^{-1/2})
\end{equation}   
where 
\begin{eqnarray}
r_{TE}(n)&=&1+\frac{1}{\pi}\int_{-\infty}^{\infty}\tilde{R}_{\mathrm{TE}}(t)\left
  ( \frac{n^2}{n^2+t^2}-\frac{1}{1+t^2}\right ) \mathrm{d}t\nonumber\\ 
&\pm &\frac{2n}{\sqrt{n^2+1}}
\label{last_term}
\end{eqnarray} 
and $\tilde{R}_{\mathrm{TE}}(t)$ is the Fresnel reflection coefficient for
the TE polarization at imaginary momentum
\begin{equation}
\tilde{R}_{\mathrm{TE}}(t)=\frac{\sqrt{n^2+t^2}-n^2\sqrt{1+t^2}}{\sqrt{n^2+t^2}+n^2\sqrt{1+t^2}}\
.
\end{equation} 
The plus-minus sign in front of the last term in \eqref{last_term} is
connected with the existence for the TE modes of an additional series of
resonances related to Brewster's angle. As these resonances have large
imaginary parts, they  may be included or not in the counting function. For
small values of $n$ additional resonances are mixed with other resonances and
their separation is artificial. In this case the plus sign has to be used.
For
large $n$ the branch of additional resonances is well separated from the body
of resonances and it is natural to ignore them. It corresponds to the minus
sign in  \eqref{last_term}.    

The results of this paper together with Ref.~\cite{trace1} demonstrate that
semiclassical trace formulas can be derived and applied for open dielectric
cavities in a close similarity with closed billiards. Further investigations
of trace formulas for other physical open systems is of considerable
interest.


\acknowledgments

It is a pleasure to thank Martin Sieber for fruitful discussions and
Stefan Bittner for providing numerical data for the dielectric circle with
$n=3$.


\appendix
\section{Krein formula approach}\label{krein}

The purpose of this Appendix is to present another  derivation of the number
of resonances in a circular dielectric cavity based on the Krein spectral
shift  formula \cite{krein_1}. The true eigen-energy spectrum for an open
system is continuous and, consequently, the density of states for open
quantum
systems is infinite. Nevertheless, the difference between the density of
states with a cavity and the density of state without the cavity is  finite
and is given by the Krein formula  
\begin{equation}
d(E)-d_0(E)=\frac{1}{2\pi \ic}\frac{\partial }{\partial E}\ln \det
\mathbf{S}(E)
\label{krein_formula}
\end{equation}    
where $\mathbf{S}(E)$ is the $S$-matrix for the scattering on the cavity. 

This formula is general and can be used for any type of short-range
potential. We apply it for a scattering on a circular dielectric cavity. It
is easy to check that the $S$-matrix for the the scattering on 2d circular
cavity with TE boundary conditions \eqref{deriv} is diagonal in the polar
coordinates and 
\begin{equation}
S_m(x)=-\frac{\tilde{s}_m(x)}{s_m(x)}
\end{equation}
where $s_m(x)$ is given by \eqref{s_m} and $\tilde{s}_m(x)$ differs from
$s_m(x)$ by changing $H_m^{(1)}(x)$ to $H_m^{(2)}(x)$:
\begin{equation}
\tilde{s}_m(x)=x\Big
[\frac{1}{n}J_m^{\prime}(nx)H_m^{(2)}(x)-J_m(nx)H_m^{(2)\,\prime}(x)\Big ]\ .
\end{equation}
From properties of the Bessel functions \cite{bateman} it is straightforward
to show that 
\begin{eqnarray}
&&\frac{s_m^{\prime}}{s_m}(x)=\\
&-&\dfrac{(n^2-1)}{n^2}\dfrac{J_m(nx)H_m^{(1)}(x)m^2/x^2+n
J_m^{\prime}(nx)H_m^{(1)\prime}(x)}{J_m^{\prime}(nx)H_m^{(1)}(x)/n- J_m(nx) 
H_m^{(1)\prime}(x)}\ .
\nonumber
\end{eqnarray}
Using the equality $J_m(x)=(H_m^{(1)}(x)+H_m^{(2)})/2$, this expression can
be rewritten in the form
\begin{equation}
\frac{s_m^{\prime}}{s_m}(x)=-\frac{n^2-1}{n^2}\dfrac{A_m(x) +B_m(x) E_m(x)
}{C_m(x)  [1-R_m(x)E_m(x) ]}
\end{equation}
where $E_m(x)$ and $R_m(x)$ are defined in \eqref{E_m} and
\eqref{exactRcircle} respectively, and 
\begin{eqnarray}
A_m(x)&=&\frac{m^2}{x^2}+n \frac{H_m^{(2)\prime}}{H_m^{(2)}}(nx)
\frac{H_m^{(1)\prime}}{H_m^{(1)}}(x)\ ,\\  
B_m(x)&=&\frac{m^2}{x^2} + n\frac{H_m^{(1)\prime}}{H_m^{(1)}}(nx)
\frac{H_m^{(1)\prime}}{H_m^{(1)}}(x)\ ,\\ 
C_m(x)&=&\frac{H_m^{(2)\prime}}{n H_m^{(2)}}(nx) -
\frac{H_m^{(1)\prime}}{H_m^{(1)}}(x)\ .
\end{eqnarray}
Expanding this expression into series of $E_m(x)$ gives
\begin{equation}
\frac{s_m^{\prime}}{s_m}(x)=Q_m(x)+P_m \sum_{r=1}^{\infty} R_m^r(x)\,
E_m^r(x)
\end{equation}
where 
\begin{equation}
Q_m(x)=-\frac{(n^2-1)A_m(x)}{n^2 C_m(x)}
\label{P}
\end{equation}
and 
\begin{equation}
P_m(x)=-\frac{4\mathrm{i} (n^2-1)\Big [\frac{m^2}{n^2 x^2}+\left
(\frac{H_m^{(1)\prime}}{H_m^{(1)}}(x) \right )^2\Big ]}
{\pi n^2 x H_m^{(1)}(nx)H_m^{(2)}(nx) C_m(x) D_m(x) } \ .
\end{equation}
with
\begin{equation}
D_m(x)=\frac{H_m^{(1)\prime}}{n
H_m^{(1)}}(nx) - \frac{H_m^{(1)\prime}}{H_m^{(1)}}(x) 
\end{equation}
In the semiclassical limit $x\to \infty$ the above formulae are simplified
by using the asymptotic of the Hankel function \eqref{hankel}
\begin{equation}
\frac{H_m^{(1,2)\prime}}{H_m^{(1,2)}}(x)\underset{x\to\infty}{\longrightarrow}
\pm \mathrm{i}\sqrt{1-\frac{m^2}{x^2}}-\frac{x}{2(x^2-m^2)}+
\mathcal{O}(x^{-2})\ .
\end{equation}
Consider first the smooth term \eqref{P}. From the identity
\begin{eqnarray}
& &\Big (\frac{1}{n^2}\sqrt{n^2-t^2}+\sqrt{1-t^2} \Big )\Big
(\sqrt{n^2-t^2}-\sqrt{1-t^2} \Big )\nonumber\\
&=&
\frac{n^2-1}{n^2}\Big (t^2+ \sqrt{n^2-t^2}\sqrt{1-t^2}  \Big )
\end{eqnarray}
it is straightforward to check that
\begin{eqnarray}
& &Q_m(x)\underset{x\to\infty}{\longrightarrow} -\mathrm{i} \left 
[\sqrt{n^2-\frac{m^2}{x^2}}-\sqrt{1-\frac{m^2}{x^2}} \right  ]\label{Q_m}\\ 
&-& \frac{x}{2} R_{\mathrm{TE}}\Big (\frac{m}{x}\Big )
\left [\frac{n^2}{n^2x^2-m^2}-\frac{1}{x^2-m^2} \right ] 
\nonumber 
\end{eqnarray}
where $R_{\mathrm{TE}}(t)$ is the Fresnel reflection coefficient for the TE
polarization given by \eqref{TE}. 

The difference between the density of state with a cavity and the one
without the cavity averaged over an  energy interval  such that periodic
orbit terms are small can be calculated from $Q_m(x)$
\begin{equation}
\langle d(E)\rangle -d_0(E)=-\frac{R}{2\pi k} \sum_{m=-\infty}^{\infty}
\mathrm{Im}\, Q_m(x)\ .
\end{equation}
Changing the summation over $m$ to the integration and  turning the
integration contour in the second term in \eqref{Q_m} in the complex plane
to avoid poles, $m\to -\ic t$ leads to
\begin{eqnarray}
&&\langle d(E)\rangle -d_0(E)=\\
&&\frac{R}{2\pi k} \Big
[\int_{-nx}^{nx}\sqrt{n^2-\frac{m^2}{x^2}}\mathrm{d}m
-\int_{-x}^{x}\sqrt{1-\frac{m^2}{x^2}}\mathrm{d}m\Big ]  \nonumber\\
&+&\frac{R x}{4\pi k} \int_{-\infty}^{\infty}\mathrm{d}t R_{\mathrm{TE}}\Big
(-\ic\frac{t}{x}\Big )
\Big [\frac{n^2}{n^2x^2+t^2}-\frac{1}{x^2+t^2} \Big ]\ .
\nonumber
\end{eqnarray} 
Rescaling integration variables one gets
\begin{eqnarray}
\langle d(E)\rangle & -& d_0(E)=\frac{\mathcal{A}}{4\pi}(n^2-1)
\label{krein_density}\\
&+&\frac{\mathcal{L} }{8\pi^2 k} \int_{-\infty}^{\infty}\mathrm{d}t \,
\tilde{R}_{\mathrm{TE}}(t)
\Big [\frac{n^2}{n^2+t^2}-\frac{1}{1+t^2} \Big ]
\nonumber
\end{eqnarray}
where $\mathcal{A}=\pi R^2$ and $\mathcal{L}=2\pi R$ are the area and the
perimeter of a circular cavity. 

This formula differs form the averaged total number of resonances
\eqref{weyl_TE} and \eqref{r_TE}. This is the consequence of the fact
discussed in \cite{trace1} for the case of TM modes that the $S$-matrix for
the scattering on a cavity has an additional phase (and additional zeros)
connected with the outside scattering on the impenetrable cavity.  

The form of this 'additional' $S$-matrix may be argued as follows. It is
known
that when a wave from outside the cavity scatters on a cavity it reflects
with
the reflection coefficient which differs by the sign from the reflection
coefficient from inside the cavity (this is a consequence of current
conservation). For the TE polarization the Fresnel reflection coefficient for
a scattering from  a medium with the refraction index $1$  on another medium
with the refraction index $n$ is $-R_{\mathrm{TE}}$ where   
$R_{\mathrm{TE}}$
is given by \eqref{TE}. In semiclassical region accessible in outside
scattering, $|t|<1$, the reflection coefficient  $-R_{\mathrm{TE}}(t)$ is
real
and the 'effective' reflection coefficient corresponding to the scattering on
impenetrable cavity equals to the sign of $-R_{\mathrm{TE}}(t)$ (cf.
\eqref{phase_delta}) 
\begin{equation}
R_{\mathrm{TE}}^{(eff)}=\left \{ \begin{array}{rr} -1, & t^*\leq |t|\leq 1\\
1,& 0\leq |t| \leq t^*\end{array}\right . 
\label{additional_reflection}
\end{equation}
where $t^*=n/\sqrt{n^2+1}$.

The reflection coefficient  equals to $-1$ (resp. $+1$)  corresponding to the
scattering with Dirichlet (resp. Neumann) conditions on the cavity boundary.
For a circular cavity the $S$-matrices with  Dirichlet and Neumann boundary
conditions are well  known (see e.g. \cite{uzy})
\begin{equation}
S_m^{(D)}(x)=-\frac{H_m^{(2)}}{H_m^{(1)}}(x),\qquad
S_m^{(N)}(x)=-\frac{H_m^{(2)\prime }}{H_m^{(1)\prime}}(x)\ .
\label{trivial_S}
\end{equation}
The 'additional' $S$-matrix for the TE scattering is thus formally 
\begin{eqnarray}
\det(S^{(\mathrm{TE})}_0(x))&=&\prod_{m=-\infty}^{-(m^*+1)}S_m^{(D)}(x)
\prod_{m=-m^*}^{m^*} S_m^{(N)}(x)\nonumber\\ 
&\times & \prod_{m=m^*+1}^{\infty}S_m^{(D)}(x) 
\label{additional_S}
\end{eqnarray}
where $m^*=[xn/ \sqrt{n^2+1}]$. 

To find the total phase of this 'additional' $S$-matrix one can proceed as
follows. To the leading order in the semiclassical limit $x\to\infty$ the
Dirichlet and Neumann $S$-matrices \eqref{trivial_S} can be calculated from
\eqref{hankel}. It gives 
\begin{equation}
S_m^{(D)}(x)\approx -\mathrm{e}^{-2\mathrm{i}\Phi_m(x)}\qquad
S_m^{(N)}(x)\approx \mathrm{e}^{-2\mathrm{i}\Phi_m(x)}
\end{equation}
where $\Phi_m(x)$ is given by \eqref{Phi}.  It means that  $S_m^{(N)}$
differs
from $S_m^{(D)}$ only by its sign, which is another manifestation of the
opposite sign of the reflection coefficient
\eqref{additional_reflection}. Therefore  one can rewrite expression
\eqref{additional_S}   as follows 
\begin{equation}
S^{(\mathrm{TE})}_0\approx S^{(D)} \prod_{m=-m^*}^{m^*}(-1)\approx S^{(D)}
\mathrm{e}^{\pm 2\mathrm{i}\pi xn/\sqrt{n^2+1} } 
\label{S_add}
\end{equation}
where $S^{(D)}= \prod_{m=-\infty}^{\infty} S_m^{(D)}$ is the full $S$-matrix
for the scattering on a cavity with the Dirichlet boundary condition.  The 
$\pm$ sign in the exponent reflects the ambiguity of the phase,
$-1=\mathrm{e}^{\pm \mathrm{i}\pi}$.
  
The calculation of the mean density of states related with  $S^{(D)}$-matrix
is straightforward (see e.g. \cite{uzy})
\begin{equation}
d_D(E)-d_0(E)=-\frac{\mathcal{A}}{4\pi}-\frac{\mathcal{L}}{8\pi k }
\end{equation}
and finally from \eqref{S_add}  and the Krein formula \eqref{krein_formula}
one finds that the change of the density of states due to the 'additional'
$S$-matrix   \eqref{additional_S} is 
\begin{eqnarray}
\bar{d}_0(E)&-&d_0(E)=d_D(E)-d_0(E) \pm \frac{\mathcal{L}}{8\pi k }
\frac{2n}{\sqrt{n^2+1}}\nonumber\\
&=& -\frac{\mathcal{A}}{4\pi}-\frac{\mathcal{L}}{8\pi k }\Big (1\mp
\frac{2n}{\sqrt{n^2+1}}\Big )\ .
\label{counter_term}
\end{eqnarray}   
The total density of resonances is thus the difference between
\eqref{krein_density} and \eqref{counter_term}. In the end one gets 
Eqs.~\eqref{weyl_TE} and \eqref{r_TE}. The ambiguity in the phase of the
'additional' $S$-matrix corresponds to the the possibility to include
resonances related with Brewster's angle in the Weyl formula or not which
has been discussed in Section~\ref{additional}.



\begin{thebibliography}{99}
\bibitem{vahala} K. Vahala, ed., \emph{Optical microcavities} (World
Scientific Press, 2004).
\bibitem{matsko} A. B. Matsko, \emph{Practical applications of microresonators
in optics and photonics}, (CRC Press, Taylor and Francis Group, 2009). 
\bibitem{balian_bloch} R. Balian, C. Bloch, Ann. Phys. \textbf{60}, 401
  (1970); Ann. Phys. \textbf{64}, 271 (1971); Ann. Phys. \textbf{69}, 76
(1972).
\bibitem{gutzwiller} M. Gutzwiller, \emph{Chaos in classical and quantum
mechanics}, (Springer-Ver\-lag, Berlin, Heidelberg, New-York, 1990). 
\bibitem{haake} F. Haake, \emph{Quantum signatures of chaos},
  (Springer-Ver\-lag, Berlin, Heidelberg, New-York, 2001).
\bibitem{scattering} P. Lax and R. S. Phillips, \emph{Scattering theory},
(Springer,  New York, 1963).
\bibitem{s_matrix} R. G. Newton, \emph{Scattering theory of waves and
particles},  (Springer-Verlag, New York, Heidelberg, Berlin, 1982). 
\bibitem{trace1} E. Bogomolny, R. Dubertrand, and C. Schmit, \pre{}  
\textbf{78}, 056202 (2008).
\bibitem{trace2} E. Bogomolny, N. Djellali, R. Dubertrand, I. Gozhyk, M.
Lebental, C. Schmit, C. Ulysse, and J. Zyss, \pre{}
  \textbf{83}, 036208 (2011). 
\bibitem{stefan2010-2012}  S. Bittner, E. Bogomolny, B. Dietz, M. Miski-Oglu,
  P. Oria Iriarte, A. Richter, F. Sch\"afer, \pre{} \textbf{81}, 066215
  (2010); S. Bittner, E. Bogomolny, B. Dietz, M. Miski-Oglu,
  A. Richter, \pre{}  \textbf{85}, 026203 (2012).
\bibitem{jackson} J. D. Jackson, \emph{Classical electrodynamics}, (Wiley,
1999).
\bibitem{melanie2007}  M. Lebental, N. Djellali, C. Arnaud,
  J.-S. Lauret, J. Zyss, R. Dubertrand, C. Schmit, E. Bogomolny, \pra{} 
\textbf{76}, 023830 (2007).
\bibitem{melanie2008} R. Dubertrand, E. Bogomolny, N. Djellali, M. Lebental,
and C. Schmit,  \pra{}  \textbf{77}, 013804 (2008).
(1997);     C. Gmachl, F. Capasso, E. E. Narimanov, J. U. N\"ockel, 
 A.~D.~Stone, J. Faist, D. L. Sivco and A. Y. Cho, Science  \textbf{280},
1556 (1998);
V. A. Podolsky, E. Narimanov, W. Fang, and H. Cao,  Proc.
Nat. Acad. Sci. USA. \textbf{101}, 10498 (2004). 
\bibitem{baltes}  H.P.  Baltes and  E.R. Hilf, \emph{Spectra of Finite
Systems}, (Bibliographisches
Institut,   Mannheim, Wien, Zurich, 1976). 
\bibitem{bateman} A. Erdelyi, \emph{Higher transcendental functions},
  Vol. II, (McGraw-Hill Book Company, New York, Toronto, London, 1955).  
\bibitem{sieber} C. P. Dettmann, G. V. Morozov, M. Sieber, and
  H. Waalkens,  Europhys. Lett. \textbf{87},  34003 (2009). 
\bibitem{krein_1} M.G. Krein,  Matem. Sbornik \textbf{33}, 597 (1953); 
Dokl. Akad. Nauk, SSR \textbf{144}, 268 (1962); English trans. in Soviet
Math. Dokl. \textbf{3} (1962).   
\bibitem{uzy} U. Smilansky and I. Ussishkin,  J. Phys. A: Math. Gen.
\textbf{29}, 2587 (1996).   
\end{thebibliography}
\end{document}